\RequirePackage{rotating}
\documentclass[12pt]{amsart}
\usepackage{graphicx}
\usepackage{amsmath,bm,amssymb}
\usepackage{color}
\newtheorem{theorem}{Theorem}[section]
\newtheorem{lemma}[theorem]{Lemma}
\newtheorem{prop}[theorem]{Proposition}
\newtheorem{cor}[theorem]{Corollary}

\theoremstyle{definition}
\newtheorem{defn}[theorem]{Definition}

\theoremstyle{remark}
\newtheorem{rem}[theorem]{Remark}

\numberwithin{equation}{section}

\newcommand{\X}{{\mathbf X}}

\newcommand{\fp}{\stackrel{p}{\rightarrow}}
\newcommand{\fd}{\stackrel{d}{\rightarrow}}
\newcommand{\eqd}{\stackrel{d}{=}}

\usepackage{array}
\usepackage{booktabs}
\usepackage{graphicx}
\usepackage{rotating}
\usepackage{tabularx}
\usepackage{float}
\graphicspath{ {./images/} }

\begin{document}
\sloppy
\title[ARTFIMA time series with stable innovations]{Statistical inference for ARTFIMA time series with stable innovations}

\author{Jinu Kabala}
\address{Jinu Kabala, Department of Statistics, Iowa State University, Ames, IA 50011}
\email{jsusan@iastate.edu}

\author{Farzad Sabzikar}
\address{Farzad Sabzikar, Department of Statistics,
Iowa State University, Ames, IA 50011}
\email{sabzikar@iastate.edu}
\urladdr{http://sabzikar.public.iastate.edu/}

\begin{abstract}
Autoregressive tempered fractionally integrated moving average with stable innovations modifies the power-law kernel of the fractionally   integrated time series model by adding an exponential tempering factor. The tempered time series is a stationary model that can exhibits semi-long-range dependence. This paper develops the basic theory of the tempered time series model, including dependence structure and parameter estimation.
\end{abstract}

\maketitle

\section{Introduction}
An autoregressive fractionally integrated moving average (ARFIMA) with stable innovations introduced in \cite{kokoszka1995fractional}. The ARFIMA model with stable innovations can be constructed using the fractional difference (or integral) operator of $\alpha$-stable innovations. The ARFIMA model is useful in practice, especially for the dataset that shows long-range dependence,
see, for example, \cite{burnecki2017identification} and the reference therein. Since the ARFIMA model with stable innovations does not have a finite second moment, then the covariance function cannot describe the dependence structures. Instead, one can use another distance measure called co-difference, see \cite{samorodnitsky1996stable}. Using this tool, Kokoszka and Taqqu investigated the asymptotic behavior of the difference of ARFIMA model with stable innovations for $\alpha\in (0,2)$, where $\alpha$ is the index of stability of a stable distribution. Later on, the parameter estimation for ARFIMA model with stable innovations studied by
\cite{kokoszka1996parameter} and \cite{burnecki2013pestimation} for stable ARTFIMA models with positive and negative memory parameter $d$, respectively.

This paper develops a model extension based on the tempered fractional difference operator introduced in \cite{meerschaert2011stochastic}. The resulting stationary time series model, called autoregressive tempered fractionally integrated moving average (ARTFIMA) with stable innovations, is obtained by replacing the fractional integral operator with a tempered fractional integral operator has additional tempering parameter $\lambda>0$.

The remainder of this paper is organized as follows. Section \ref{sec:ARTFIMAdef} develops the stable ARTFIMA model, proves causality and invertibility. Unlike the ARFIMA model, the stable ARTFIMA model is stationary and invertible for any $d\in\mathbb{R}-{Z}_{-}$ due to the presence of an exponential tempering function in its moving average representation. Section \ref{dependeceARTFIMAstable} computes the dependence structure of stable ARTFIMA, which can exhibit the heavy-tailed analog of semi-long range dependence. Section \ref{sec:parameterestimation} presents the parameter estimation of stable ARTFIMA model, including the estimators' consistency and asymptotic distributions. Sections \ref{simulations} and \ref{Solar} contain the simulation of stable ARTFIMA model and an application of this model to Solar-Flare dataset respectively. All the proofs of the paper results are collected in the appendix \ref{proofs}.

\section{The ARTFIMA model with stable innovations}\label{sec:ARTFIMAdef}
This section defines the ARTFIMA model with stable innovations and studies its essential properties, including causality and invertibility. First, we provide some basic definitions and assumptions.

The tempered fractionally integrated operator is defined by:
\begin{equation}\label{eq:TFdiffDef}
\Delta^{-d,\lambda} f(x)=(I-e^{-\lambda }B)^{-d} f(x)=\sum_{j=0}^\infty \omega_{-d,\lambda}(j)f(x-j),
\end{equation}
where $d$ is a positive non-integer value, the tempering parameter $\lambda>0$, $BX(t) = X(t-1)$ is the shift operator, and
\begin{equation}\label{omegadef}
\omega_{-d,\lambda}(j):=\frac{\Gamma(j+d)}{\Gamma(d)\Gamma(j+1)} e^{-\lambda j}
\end{equation}
with the gamma function $\Gamma(d)=\int_0^\infty e^{-x}x^{d-1} dx$. Applying the well-known property $\Gamma(d+1)=d\Gamma(d)$, we can extend \eqref{eq:TFdiffDef} to non-integer values of $d<0$. In this case, $\Delta^{-d,\lambda}$ is called tempered fractional difference operator with the positive exponent $-d$. If $\lambda=0$, then equation \eqref{eq:TFdiffDef} reduces to the usual fractionally integrated operator. See \cite{sabzikar2015TFC,meerschaert2014TFI} for more details.

A $S\alpha S$ random variable $Z_{\alpha}=\{Z_{\alpha}(t)\}$ has characteristic function
\begin{equation}
\mathbb{E}[e^{i\theta Z_{\alpha}(t)}]=e^{-\sigma^{\alpha}|\theta|^{\alpha}|t|}, \quad \theta\in\mathbb{R}, 0<\alpha\leq 2.
\end{equation}
The parameter $\sigma$ is called the scale parameter of $Z$. If $\alpha=2$, $Z$ is Gaussian with
variance $2\sigma^2$~. If $0 < \alpha < 2$, then $\mathbb{E}[|Z|^{p}]=\infty$ for $p \geq \alpha$, and for $0 < p < \alpha$
\begin{equation}\label{pthmoment}
\mathbb{E}[|Z|^{p}] = \mathbb{E}[|\sigma^{-1}Z|^{p}]\sigma^{p}= c(p, a)\sigma^{p},
\end{equation}
where the constant $c(p,\alpha)$ does not depend on the scale parameter $\sigma$.

Next, we provide two main assumptions that we will use for the rest of the paper.

{\bf Assumption 1:} In this paper, we assume $\{Z(t)\}$ is a sequence of i.i.d $S\alpha S$ random variables with the scale parameter $\sigma=1$.

{\bf Assumption 2:} Let $\Phi_p(z)$ and $\Theta_q(z)$ be polynomial with real coefficients defined by
 $\Phi_p(z)=1-\phi_1 z-\phi_2 z^2-\ldots -\phi_{p}z^{p}$, and $\Theta_q(z)=1+\theta_1 z+\theta_2 z^2+\ldots +\theta_{q}z^{q}$ respectively. In this paper, we always assume the polynomials $\Phi_p$ and $ \Theta_q$ have no common roots and the polynomial $\Phi_p$ has no roots in the closed unit disk $\{z:|z|\leq 1 \}$.

\begin{defn}\label{def:artf}
The stochastic process $\{X(t)\}_{t\in\mathbb{Z}}$ is said to be stable {\it autoregressive tempered fractional integrated moving average}, denoted by ARTFIMA$(p,d,\lambda,q)$, model if $\{X(t)\}_{t\in Z}$ satisfies the tempered fractional difference equations
\begin{equation*}
\Phi(B)X(t) = \Theta(B)\Delta^{-d,\lambda} Z(t),
\end{equation*}
where $\{Z(t)\}_{t\in\mathbb{Z}}$ are i.i.d $S\alpha S$ for $0<\alpha\leq 2$, $\Phi_p, \Theta_q$ are polynomials in Assumption 2, $d\in \mathbb{R}-\{-1,-2,\cdots\}$, and $\lambda>0$.
\end{defn}

The next proposition shows the ARTFIMA$(p,d,\lambda,q)$ with stable innovations is causal and invertible.

\begin{prop}\label{th:marep}
Suppose that $\{X(t)\}_{t\in\mathbb{Z}}$ is an ARTFIMA$(p,d,\lambda,q)$ time series that satisfies Definition \ref{def:artf}. Then,
\begin{itemize}
\item [(a)] $X(t)$ has the moving average representation
\begin{equation}\label{movingaveragerep}
X(t)=X_{p,d,\lambda,q}(t)=\sum_{j=0}^{\infty} a_{-d,\lambda}(j)Z(t-j),
\end{equation}
where
\begin{equation*}\label{acoefficient}
a_{-d,\lambda}(j)=\sum_{s=0}^{j}\omega_{-d,\lambda}(s)b(j-s)
\end{equation*}
with $\Theta_q(z)\Phi_{p}(z)^{-1}=\sum_{j=0}^{\infty}b(j)z^j$ for $|z|\leq 1$. Moreover, the series in \eqref{movingaveragerep} converges a.s. and in $L^{\nu}$ for any $\nu<\alpha$.
\item [(b)] $X(t)$ is invertible. That is
\begin{equation}\label{invertrep}
Z(t)= \sum_{j=0}^{\infty}c_{d,\lambda}(j)X(t-j),
\end{equation}
where
\begin{equation*}\label{ccoefficient}
c_{d,\lambda}(j)=\sum_{s=0}^{j}\omega_{d,\lambda}(j)c(j-s)
\end{equation*}
with $\frac{\Phi_{p}(z)}{\Theta_q(z)}=\sum_{j=0}^{\infty}c(j)z^j$ for $|z|\leq 1$. Moreover, the series in \eqref{invertrep} converges a.s. and in $L^{\nu}$ for any $\nu<\alpha$.
\end{itemize}
\end{prop}

\begin{rem}
\begin{itemize}
\item [(a)] There is another version of an ARTFIMA$(p,d,\lambda,q)$ that was introduced in \cite{sabzikar2018} in the following sense: A discrete time stochastic process $X^{I\!I}_{p,d,\lambda,q}$ is called ARTFIMA$(p,d,\lambda,q)$ process with stable innovations $\{Z(t)\}$ if
\begin{equation}\label{eq:Xjdefinition}
X^{I\!I}_{p,d,\lambda,q}(t) = \sum_{j=0}^{\infty} e^{-\lambda j}a_{-d}(j)  Z(t-k), \qquad t \in \mathbb{Z},
\end{equation}
where
\begin{equation*}\label{acoef}
a_{-d}(j) = \sum_{s=0}^{j} \omega_{-d}(s) b(j-s), \qquad k \ge 0,
\end{equation*}
where  $\omega_{-d}(j) = \omega_{-d,0}(j) =\frac{\Gamma(k+d)}{\Gamma(k+1)\Gamma(d)}$.
The stochastic process $\{X^{I\!I}_{p,d,\lambda,q}\}$ should be considered as ARTFIMA$(p,d,\lambda,q)$ of the second kind.
\item [(b)] We note that $\{X_{p,d,\lambda,q}\}$ in \eqref{movingaveragerep} and $\{X^{I\!I}_{p,d,\lambda,q}\}$ in \eqref{eq:Xjdefinition} are two different stochastic processes since $a_{-d,\lambda}(j)\neq e^{-\lambda j}a_{-d}(j)$ in general. However, these two processes are the same if $\Phi_{p}(z)=\Theta_{q}(z)=1$ which means $X_{0,d,\lambda,0}=X^{I\!I}_{0,d,\lambda,0}$. In this paper, we focus on the $X_{p,d,\lambda,q}$.
\end{itemize}
\end{rem}

\section{Dependence structure of ARTFIMA model with stable innovations}\label{dependeceARTFIMAstable}
In this section, we investigate the dependence structure of the ARTFIMA model with stable innovations. The covariance function stops working in the presence of stable innovations since the finite second moment does not exist. Instead, we will use other distance measure which is called co-difference to describe the dependence structure. Let $X(n)=\sum_{j=0}^{\infty}c(j) Z(n-j)$ be a casual moving average with
$S\alpha S$ innovations $\{Z(t)\}$ for $\alpha\in (0,2)$. A co-difference $\tau_{n}$ of $X_0$ and $X(n)$ is defined by
\begin{equation*}\label{codifferencedefinition}
\tau_{n}:=\tau(X(0),X(n))=\sum_{j=0}^{\infty}\big[|c(j)|^{\alpha}+|c(j+n)|^{\alpha}-|c(j) - c(j+n)|^{\alpha}\big].
\end{equation*}
Note that if $\alpha=2$, then $\tau(X(0),X(n))={\rm cov}(X(0),X(n))$. We refer the reader to \cite{samorodnitsky1996stable} for more properties of the co-difference.

The next two theorems provide the asymptotic behavior of the stable ARTFIMA model's co-difference when $\alpha\in (0,1)$ and $\alpha\in (1,2)$ respectively. Without loss of generality, we may consider stable ARTFIMA$(0,d,\lambda,0)$ in the next two Theorems.

\begin{theorem}\label{codifferenceless1}
Let $X_{d,\lambda}(t)=\sum_{j=0}^{\infty} \omega_{-d,\lambda}(j) Z_{t-j}$ be an ARTFIMA$(0,d,\lambda,0)$ with $S\alpha S$ innovations $\{Z(t)\}$. Suppose $\alpha\in (0,1)$, $d\in\mathbb{R}-\mathbb{N}_{-}$, and $\lambda>0$. Then
\begin{equation*}\label{codifferencelambdan}
\lim_{n\to\infty}\frac{\tau_{d,\lambda}(n)}{e^{-\lambda\alpha n}n^{\alpha(d-1)}} =(\Gamma(d))^{-\alpha}(1-e^{-\lambda \alpha})^{-1}.
\end{equation*}
\end{theorem}

\begin{theorem}\label{codifferencegrq1}
Let $X_{d,\lambda}(t)=\sum_{j=0}^{\infty} a_{-d,\lambda}(j) Z_{t-j}$ be an ARTFIMA$(0,d,\lambda,0)$ with $S\alpha S$ innovations $\{Z(t)\}$. Suppose $\alpha\in (1,2)$, $d\in\mathbb{R}-\mathbb{N}_{-}$, and $\lambda>0$. Then
\begin{equation*}\label{codifferencelambdan}
\lim_{n\to\infty}\frac{\tau_{d,\lambda}(n)}{e^{-\lambda n}n^{d-1}} = \frac{\alpha}{\Gamma(d)}\sum_{j=0}^{\infty} e^{-\lambda j} \omega_{-d,\lambda}^{\alpha-1}(j).
\end{equation*}
\end{theorem}

A stationary casual moving average representation $X(n)$ with finite second moments innovations, i.e. $\mathbb{E}(Z^2)<\infty$ is called to have long memory if $\sum_{n=0}^{\infty}\gamma(n)=\infty$ where $\gamma(n)={\rm Cov}(X(0),X(n))$. In the absence of finite second moment innovation, i.e. $\mathbb{E}(Z^2)=\infty$, $X(n)$ is called to have long memory if
\begin{equation}\label{longmemorystable}
\sum_{n=0}^{\infty}|\tau(n)|=\infty
\end{equation}
where $\tau(n)$ is the co-difference defined by \eqref{codifferencedefinition}.

\begin{cor}\label{semilong}
The stable ARTFIMA$(0,d,\lambda,0)$ does not have long memory in the sense of \eqref{longmemorystable}.
\end{cor}

\begin{figure}[htpb]
\includegraphics[width=2.9in]{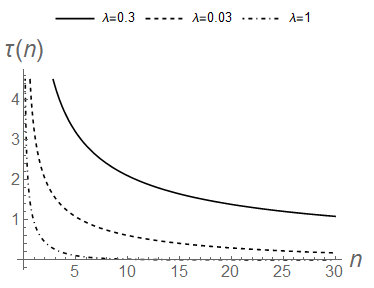}
\includegraphics[width=3in]{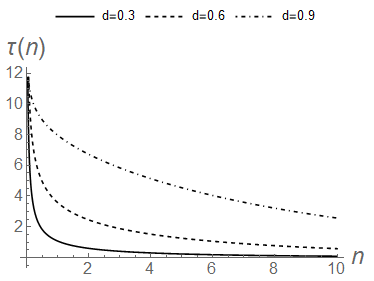}
\caption{Theoretical co-difference function for ARTFIMA$(0,d,\lambda,0)$ model. The left panel shows the co-difference for $d=0.4$ and different values of $\lambda$. The right panel shows the codifference for $\lambda=0.1$ and different values of $d$.}\label{fig1}
\end{figure}

\begin{rem}
According to Corollary \ref{semilong}, the stable ARTFIMA model is not long-range dependent. But, it does exhibit {\it semi-long range dependence} under the assumptions of Theorems \ref{codifferenceless1} and \ref{codifferencegrq1}. That is, for a sufficiently small value of $\lambda$, the sum in \eqref{longmemorystable} is large because it tends to infinity as $\lambda$ approaches to zero. Figure \ref{fig1} illustrates the co-difference of the ARTFIMA$(0, d, \lambda, 0)$
model. The differencing parameter $d$ behaves similarly to that of ARFIMA
model. A higher value of $d$ gives a stronger co-difference, so the co-difference falls off more
slowly with lag. A higher value of the tempering parameter $\lambda$ makes the co-difference fall off
more rapidly.
\end{rem}

\begin{rem}
The explicit form and asymptotic behavior of the covariance function for stationary ARTFIMA $X^{I\!I}_{0,d,\lambda,0}$ when the innovations have the finite second moment obtained in \cite{sabzikar2018}. In this case,
$\rm Cov(X^{I\!I}_{0,d,\lambda,0}(0)X^{I\!I}_{0,d,\lambda,0}(k))\sim C k^{d-1}e^{-\lambda k}$ as $k\to\infty$. By Letting $\Phi_p=\Theta_q=0$ and the innovations $Z(t)$ have the finite second moment in ARTFIMA$(0,d,\lambda,0)$, Theorem \ref{codifferencegrq1} implies that
$\tau_{d,\lambda}(n)\sim C e^{-\lambda n}n^{d-1}$ which is similar to the results in \cite{sabzikar2018}. In other words, when $1<\alpha\leq 2$, the power law and exponential function involved in the asymptotic behavior of the co-difference and covariance function does not depend on $\alpha$.
\end{rem}

%
%
%

%


\section{Parameter Estimation}\label{sec:parameterestimation}
In this section, we prove the consistency and asymptotic distribution of a variant Whittle estimator for ARTFIMA$(p,d,\lambda,q)$. The Whittle's method for ARMA$(p,q)$ model was studied by Mikosch et. al \cite{mikosch1995parameter}. An extension of this method was developed by Kokoszka and Taqqu \cite{kokoszka1996parameter} for ARFIMA$(p,d,q)$ when $d\in(0,1/2)$ and $\alpha\in (1,2)$. Burnecki and Sikora \cite{burnecki2013pestimation} investigated the parameter estimation for ARFIMA$(p,d,q)$ when $d\in(-1/2,0)$ and $\alpha\in (2/3,2]$. In all these references, there is a function which is called {\it power transfer function} and it plays an important role to show the consistency of the parameters. Following the Whittle's method in  \cite{mikosch1995parameter,kokoszka1996parameter,burnecki2013pestimation}, we modify the power transfer function to our case as follows. We define the {\it tempered power transform function}
\begin{equation}\label{transfer}
g_{p,d,\lambda,q}(\omega,{{\bf \beta}}):=\Big|\frac{\Theta_{q}(e^{-i\omega}, {\bf \beta})}{\Phi_{p}(e^{-i\omega}, {\bf \beta})(1-e^{-(\lambda+i\omega)})^{d}} \Big|^{2},
\end{equation}
where the $(p+q+2)$ dimensional vector ${\bf\beta}=(\phi_1, \cdots, \phi_p, d, \lambda, \theta_1, \cdots, \theta_q)$ belongs to the parameter space
\begin{equation}\label{parameter_space}
\begin{split}
E:&=\Big\{{\bf \beta}: \phi_p, \theta_q\neq 0, \Phi_p(z), \Theta_q(z)\neq 0\ {\rm for}\ |z|\leq 1,  \\
&\ d\in \mathbb{R}-\mathbb{Z}_{-}, \lambda>0 \Big\}.
\end{split}
\end{equation}
We should note that $g_{p,0,0,q}$ and $g_{p,d,0,q}$ reduce to the transfer function in \cite{mikosch1995parameter} and
\cite{kokoszka1996parameter,burnecki2013pestimation} respectively.

Next, to establish consistency and asymptotic distribution of our estimators, we set up some assumptions on the innovations $\{Z(t)\}$ of the stable ARTFIMA$(p,d,\lambda,q)$ given by \eqref{movingaveragerep}:

{\bf Assumption 3:} The innovations $\{Z(t)\}$ satisfy under the following three assumptions:
\begin{itemize}
\item [(3-a)] $E|Z_1|^\nu < \infty$,\ {\rm for\ some}\  $\nu>0$.
\item [(3-b)] $n^{1-{2\delta/\alpha}}\to 0$,\ {\rm as}\ $n\to\infty$, \ {\rm for}\ $\delta = 1\land \nu$.
\item [(3-c)] $\lim_{x\to 0}\limsup_{n\to\infty} \mathbb{P}\Big(n^{-2/\alpha}\sum_{t=1}^n Z^{2}(t)\leq x\Big)=0$
\end{itemize}

{\bf Assumption 4:} The innovations $\{Z(t)\}$ belong to the domain of normal attraction (DNA) of a symmetric $\alpha$-stable, random variable $Z(t)\in {\rm DNA}(\alpha)$ for some $0<\alpha<2$. That is  $\frac{1}{n^{1/\alpha}}\sum_{t=1}^{n}Z(t) \Rightarrow^d Y$,
where $Y$  is symmetric $\alpha$-stable.

Let $\X=(X(1),\ldots,X(n))$ be a realization of the stable ARTFIMA$(p,d,\lambda,q)$ time series with sample size $n$ and define the self-normalised periodogram
\begin{equation}\label{eq:selfnormperiodogram definition}
\tilde{I}_{\X}(\omega):=\frac{\Big|\sum_{t=1}^{n}X(t)e^{-it\omega}\Big|^2}{\sum_{t=1}^{n}X^{2}(t)},\quad -\pi<\omega<\pi.
\end{equation}
Let $\beta=(\phi_1,\ldots, \phi_p,d,\lambda,\theta_1, \ldots, \theta_q)$ be a vector in the parameter space $E$ in \eqref{parameter_space}. Define
\begin{equation}\label{sigmaestimator}
\sigma_n^2 (\beta) =\int_{-\pi}^{\pi}\frac{\tilde{I}_{\X}(\omega)}{g_{p,d,\lambda,q}(\omega)}\ d\omega,
\end{equation}
where $g_{p,d,\lambda,q}$, $\tilde{I}_{\X}$ are given by \eqref{transfer} and \eqref{eq:selfnormperiodogram definition} respectively.

\begin{defn}
Let $\beta_{0}$ denote the true parameter values of $\beta$. The {\it estimators} of $\beta_{0}$ based on $\X=(X(1),\ldots,X(n))$ are defined by \eqref{parameter_space}.
\begin{equation*}\label{def:WhittleEst}
\beta_n:={\arg\min} \{\sigma_n^2 (\beta):\beta\in E\},
\end{equation*}
where $E$ is the parameter space given by
\end{defn}

\begin{lemma}\label{lemma1consistency}
Let $\{X(t)\}_{t\in\mathbb{Z}}$ be the $ARTFIMA(p,d,\lambda,q)$ given by \eqref{movingaveragerep} with innovations $\{Z(t)\}$ hold on condition (3-a) of assumption 3. Then,
\begin{equation*}
\sum_{j=-\infty}^\infty |a_{-d,\lambda}(j)|^\delta |j| < \infty
\end{equation*}
for $\delta = 1\land \eta$.
\end{lemma}

Define the periodogram for $X(t)$ as
\[I_{X}(\omega)=n^{-2/\alpha}\Big|\sum_{t=1}^n X(t) e^{-i\omega t}\Big|, \indent -\pi<\omega\leq \pi\]
and let $I_{Z}(\omega)$ be the periodogram of the innovations $Z(t)$. For any $h\in\mathbb{Z}$ define
\begin{equation*}
\tilde{\gamma}_{n,X}(h)= \frac{\gamma_{n,X}(h)}{\gamma^2_{n,X}},\ \tilde{\gamma}(h)=\frac{\gamma(h)}{\gamma(0)},
\end{equation*}
where
 \begin{eqnarray*}
  \gamma_{n,X}(h) = n^{-2/\alpha}\sum_{t=1}^{n-|h|}X(t) X(t+|h|) &, h\in\mathbb{Z},\\
  \gamma(h)=\sum_{h=-\infty}^\infty a_{-d,\lambda}(j)a_{-d,\lambda}(j+|h|) &, h\in\mathbb{Z},\\
  \gamma^2_{n,X}= n^{-2/\alpha} \sum_{t=1}^n X^{2}(t).
\end{eqnarray*}
The quantities  $\gamma^{2}_{n,Z}, \gamma_{n,Z}(h)$ and $\tilde{\gamma}_{n,Z}(h)$ can be defined similarly.

\begin{theorem}\label{thm:asymptotic_dis_covariance}
Let $\{X(t)\}_{t\in\mathbb{Z}}$ is a stable $ARTFIMA(p,d,\lambda,q)$ with the innovations $\{Z(t)\}$ satisfying
Assumptions 3 and 4. Then for any positive integer $h$,
\begin{equation*}
\Big(\frac{n}{\ln n}\Big)^{1/\alpha}
\Big(\tilde{\gamma}_{n,X}(1)-\gamma_{n,X}(1), \cdots, \tilde{\gamma}_{n,X}(h)-\gamma_{n,X}(h) \Big)^{\prime}\fd
(S(1), \cdots, S(h))^{\prime},
\end{equation*}
where
\begin{equation*}
S(k)= \sum_{j=1}^{\infty}\Big(\gamma_{n,X}(k+j)+\gamma_{n,X}(k-j)-2\gamma_{n,X}(j)\gamma_{n,X}(k)                                        \Big)\frac{Y(j)}{Y(0)},\ k=1,\cdots, h,
\end{equation*}
and $Y(0), Y(1),\cdots, Y(k)$ are independent symmetric stable random variables. Moreover,
$Y(0)\eqd S_{\alpha/2}(C^{-2}_{\alpha/2},1,0)$ is a positive $\alpha/2$-stable, $\{Y(k)\}_{k\geq 1}$ are i.i.d. $S\alpha S$
with scale parameter $\sigma= C^{-1/\alpha}_{\alpha}$, where
\begin{equation*}
C_{\alpha}=\begin{cases}
\frac{1-\alpha}{\Gamma(2-\alpha)\cos(\pi\alpha/2)}, &\alpha\neq 1, \\
\frac{2}{\pi},&  \alpha=1.
\end{cases}
\end{equation*}
\end{theorem}

\begin{lemma}\label{lemma2consistency}
Let $\{X(t)\}_{t\in\mathbb{Z}}$ be a stable $ARTFIMA(p,d,\lambda,q)$ with the innovations $\{Z(t)\}$ satisfying
assumption 3. Then
\begin{itemize}
\item [(1)] \begin{equation}\label{part1_lemma2consistency}
\sigma_n^2(\beta) \fp \frac{1}{\gamma(0)} \int_{-\pi}^{\pi} \frac{g_{p_0,d_0,\lambda_0,q_0}(\omega,\beta_0)}
{g_{p,d,\lambda,q}(\omega,\beta)}d\omega,
\end{equation}
where $\beta_0=(\phi_1, \ldots, \phi_{p_0},d_0,\lambda_0,\theta_1,\ldots, \theta_{q_0} )$ is the true unknown parameter in $E$ and
$\gamma(0)=\sum_{j=0}^{\infty}a^{2}_{-d,\lambda}(0)$.
\item [(2)] For every $\delta > 0$,
\begin{equation}\label{part2_lemma2consistency}
\sup_{\beta\in {\bar E}}\Big|\sigma_{n,\delta}^2(\beta) - \frac{1}{\gamma(0)}
\int_{-\pi}^{\pi} \frac{g_{p_0,d_0,\lambda_0,q_0}(\omega,\beta_0)}
{g^{\delta}_{p,d,\lambda,q}(\omega,\beta)}d\omega\Big| \fp 0,
\end{equation}
where $\bar{E}$ denotes the closure of $E$,
\[g^{\delta}_{p,d,\lambda,q}(\omega,\beta) := \left(\frac{|\theta(e^{-i\nu})|^2 + \delta}{|\phi(e^{-i\nu})|^2}\right)(1-2e^{-\lambda} \cos \nu + e^{-2\lambda})^{-d},\]
and
\[\sigma^{2}_{n,\delta}(\beta) = \int_{-\pi}^{\pi} \frac{\tilde{I}_{\X}(\omega
)}{g^{\delta}_{p,d,\lambda,q}(\omega,\beta) d\omega}.\]
\end{itemize}
\end{lemma}

\begin{lemma}\label{lemma3consistency}
Suppose $\beta_{i}=(\phi_1, \ldots, \phi_{p_i}, d_i, \lambda_i, \theta_1, \cdots, \theta_{q_i})\in E$ for $i=1,2$. If $\beta_1\neq \beta_2$, then
\begin{equation*}
\frac{1}{2\pi}\int_{-\pi}^{\pi}\frac{g_{p_1,d_1,\lambda_1,q_1}(\omega,\beta_1)}{g_{p_2,d_2,\lambda_2,q_2}(\omega,\beta_2)}>1.
\end{equation*}
\end{lemma}

\begin{theorem}\label{thm:consistency}
Let $\{X(t)\}_{t\in\mathbb{Z}}$ is a stable $ARTFIMA(p,d,\lambda,q)$ with the innovations $\{Z(t)\}$ satisfying
Assumption 3. Then
\begin{equation*}
\beta_n\fp \beta_0\ {\rm and}\ \sigma_n^2(\beta_n) \fp \frac{2\pi}{\gamma(0)}.
\end{equation*}
\end{theorem}

\begin{theorem}\label{thm:asymptotic_dis_parameters}
Let $\{X(t)\}_{t\in\mathbb{Z}}$ is a stable $ARTFIMA(p,d,\lambda,q)$ with the innovations $\{Z(t)\}$ satisfying
Assumption 4. Then
\begin{equation*}
\Big(\frac{n}{\ln n}\Big)^{1/\alpha}\big(\beta_n-\beta_0\big)\fp 4\pi {\bf W}^{-1}(\beta_0)\frac{1}{Y(0)}\sum_{k=1}^{\infty}Y(k) b(k),
\end{equation*}
where
$Y(0)\eqd S_{\alpha/2}(C^{-2}_{\alpha/2},1,0)$ is a positive $\alpha/2$-stable, $\{Y(t)\}_{t\in \mathbb{N}}$ are i.i.d. $S\alpha S$
with scale parameter $\sigma= C^{-1/\alpha}_{\alpha}$, ${\bf W}^{-1}(\beta_0)$ is the inverse of the matrix
\begin{equation*}\label{eq:Wmatrix}
{\bf W}(\beta_0)= \int_{-\pi}^{\pi}\Big\{\frac{\partial\log g_{p,d,\lambda,q}(\omega,\beta_0)}{\partial{\beta}}\Big\}\Big\{\frac{\partial\log g_{p,d,\lambda,q}(\omega,\beta_0)}{\partial{\beta}}\Big\}^\prime d\omega
\end{equation*}
and
\begin{equation*}
b(k)=\frac{1}{2\pi}\int_{-\pi}^{\pi} e^{-i k\omega}g_{p,d,\lambda,q}(\omega,\beta_0)
\Big( \frac{\partial g^{-1}_{p,d,\lambda,q}(\omega,\beta_0)}{\partial{\beta}}\Big)\ d\omega,
\end{equation*}
where $g^{-1}_{p,d,\lambda,q}$ is the reciprocal function of $g_{p,d,\lambda,q}$.
\end{theorem}

\section{Simulation Results}\label{simulations}
\begin{figure}[htp]
\includegraphics[width=0.48\linewidth]{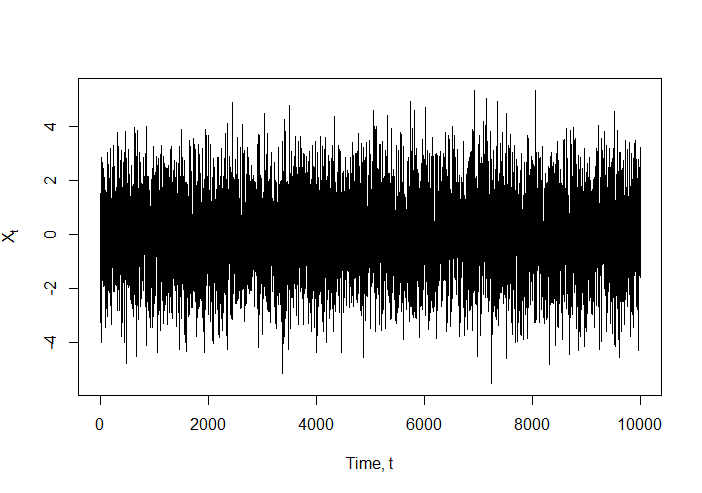}
\includegraphics[width=0.48\linewidth]{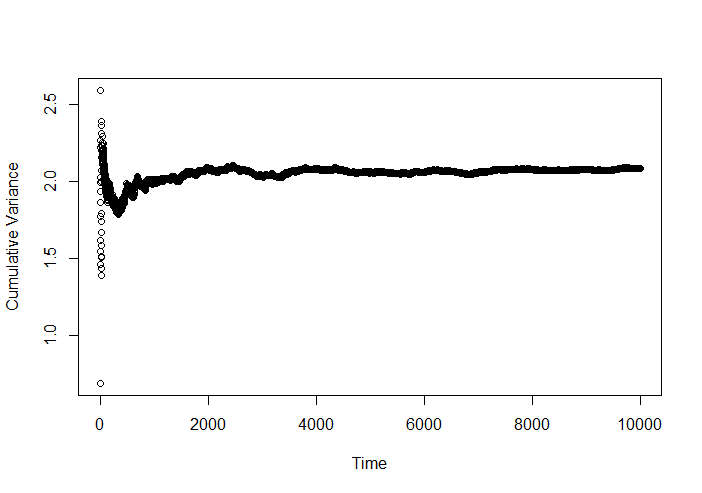}
\caption{Plot of 10000 samples of an ARTFIMA time series (left) with stable innovations $\alpha=2, d=0.1, \lambda=0.045$ along with a cumulative variance plot (right) showing converging variance due to Gaussian distribution.}
\label{fig:simres_alpha2.0}
\end{figure}

\begin{figure}[htp]
\includegraphics[width=0.48\linewidth]{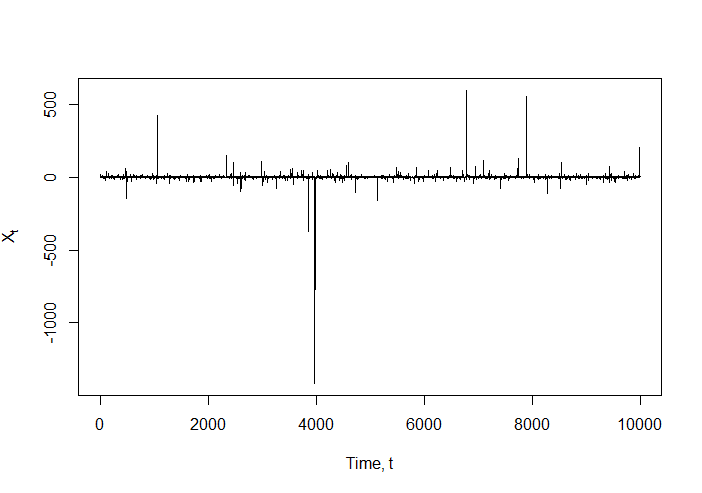}
\includegraphics[width=0.48\linewidth]{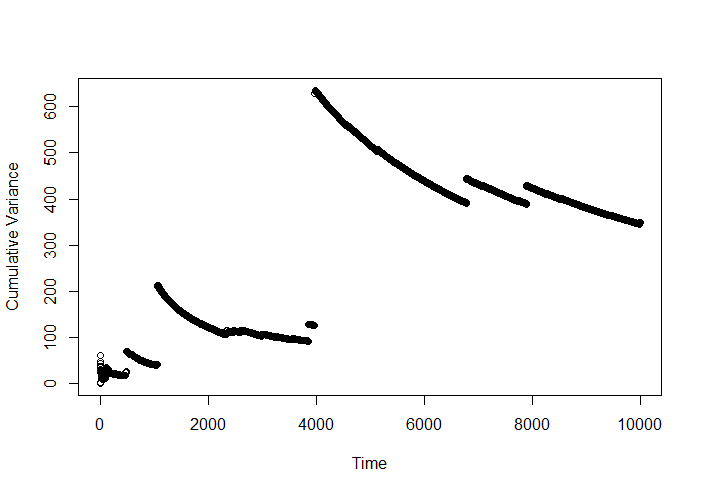}
\caption{Plot of 10000 samples of an ARTFIMA time series (left) with stable innovations $\alpha=1.3, d=0.1, \lambda=0.045$ along with a cumulative variance plot (right) showing diverging variance due to stable distribution.}
\label{fig:simres_alpha1.3}
\end{figure}

\begin{figure}[htp]
\includegraphics[width=0.48\linewidth]{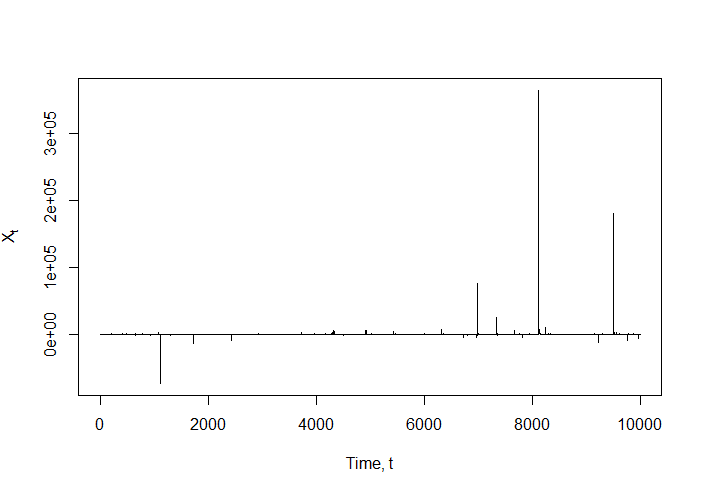}
\includegraphics[width=0.48\linewidth]{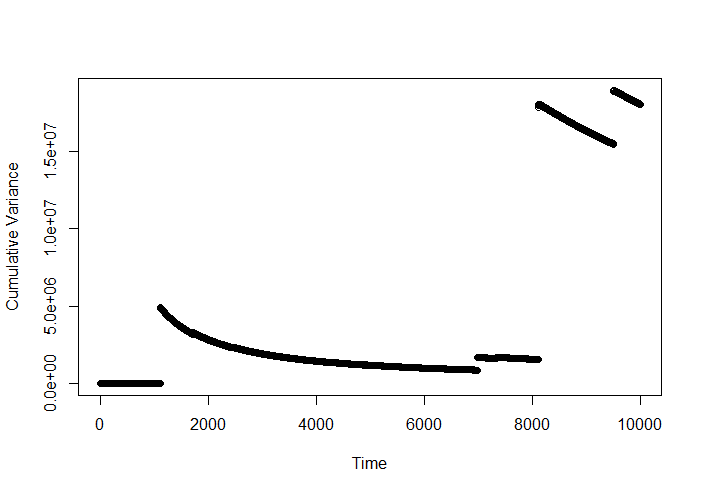}
\caption{Plot of 10000 samples of an ARTFIMA time series (left) with stable innovations $\alpha=0.7, d=0.1, \lambda=0.045$ along with a cumulative variance plot (right) showing diverging variance due to stable distribution.}
\label{fig:simres_alpha0.7}
\end{figure}

In this section, we simulate ARTFIMA model with stable innovations. In order to simulate stable ARTFIMA time series, we use the Durbin-Levinson algorithm \cite{kokoszka2001can} in which the innovations are replaced by the stable simulations \cite{chambers1976method}.
We consider ARTFIMA$(0,0.3,0.045,0)$ of length 10000 when $\alpha \in \{2.0, 1.3, 0.7\}$. The time series plot and the corresponding cumulative variance plot that illustrates infinite variance \cite{adler1998analysing} are shown in Figures \ref{fig:simres_alpha2.0}, \ref{fig:simres_alpha1.3} and \ref{fig:simres_alpha0.7}.  The Gaussian time series when $\alpha=2$ lacks the jumps in the cumulative variance plot which is evident when $\alpha \in \{1.3,0.7\}$ where the variance diverges with time.

\begin{table}[htp]
\centering
\caption{Whittle estimates of $d, \lambda$ for different stable time series along with average, bias, mean square error and 95\% percentile bootstrap confidence intervals of the estimates.}
\label{tab:simu_paramest}
\resizebox{\linewidth}{!}{\begin{tabular}{c|cccc|cccc|}
\toprule
& \multicolumn{4}{c|}{$d= 0.1$}& \multicolumn{4}{c|}{$\lambda=0.045$}\\
\cmidrule{2-9}
 {$\alpha$}&${\rm Mean}$  & ${\rm Bias}$    & ${\rm MSE}$   &  ${\rm CI}$ & ${\rm Mean}$  & ${\rm Bias}$    & ${\rm MSE}$   &  ${\rm CI}$\\
\hline
 $2$  & 0.102 & 0.002 & 1.56e-07 & [0.081, 0.13] & 0.062 & 0.0176 & 2.878e-06 & [0.002, 0.201]\\
 $1.3$ & 0.104 & 0.004 & 5.0e-06 & [0.088, 0.121] & 0.059 & 0.014 & 1.30e-05 & [0.003,  0.151]\\
  $0.7$  & 0.103 & 0.003 & 6.11e-06 & [0.093, 0.107] & 0.054 & 0.009 & 1.42e-05 & [0.005, 0.086]\\
\bottomrule
\end{tabular}}
\end{table}

For the time series shown in Figures \ref{fig:simres_alpha2.0}, \ref{fig:simres_alpha1.3} and \ref{fig:simres_alpha0.7}, we compute the whittle estimates of parameters $d$ and $\lambda$ for 1000 Monte-carlo simulations of the time series. We calculate the bias, mean square error (MSE), and 95\% percentile bootstrap confidence interval for the time series as shown in Table \ref{tab:simu_paramest}. The parameter estimates exhibit very low bias and MSE. The $95\%$ bootstrap confidence interval and the bias for parameter $\lambda$ is larger than that of $d$.

\section{An Application to Solar-flare data}\label{Solar}

\begin{figure*}[!ht]
\centering
\includegraphics[width=0.5\linewidth]{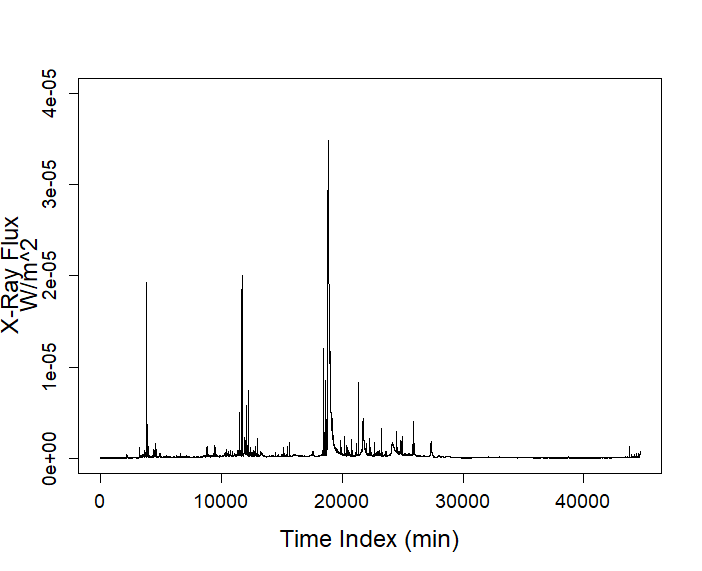}
\caption{Time series plot of the Solar flare energy for July 2017}
\label{fig:avgsolar_alphad_tsplot}
\end{figure*}

The solar flare event data is collected from the X-ray sensors (XRS) on the GOES \cite{goes8solarflaredata} satellites provided by the NOAA Space Weather Prediction Center (SWPC). The data consists of X-ray fluxes from two channels: a short channel (wavelength bands of 0.5 to 4  $\AA$) and a long channel (wavelength 1 to 8  $\AA$). These XRS channels are prone to saturate under extreme flare events. To get the true fluxes, the SWPC scaling factors are removed by dividing the XRS flux from the long channel by 0.7 \cite{solarflaredatareadme}. The two bands of X-rays: 1-8 $\AA$ band and 0.5-4 $\AA$ band are also called soft and hard X-ray emission respectively. In this paper, we use X-ray fluxes from the long channel or soft X-ray emissions. For July, data is available from both GOES-13 and GOES-15 satellites. For those time points for which multiple readings exist, we take the average of the available readings; else, we take the maximum of the readings. The resulting time series for July 2017 is of length 44640, out of which 44416 values are an average of both GOES-13, and GOES-15 readings, and the remaining 224 values are the maximum of the reading. The July 2017 time series shown in Figure \ref{fig:avgsolar_alphad_tsplot}. 
As described in \cite{stanislavsky2019solar}, we use a Hidden Markov Model to extract shorter stationary trajectories from time series, which follow a stable distribution. The sub-series between indexes 7400 to 8400 of the July 2017 time series is the dataset that we use in the remaining sections.

\begin{figure*}[htpb]
\centering
\includegraphics[width=0.5\linewidth]{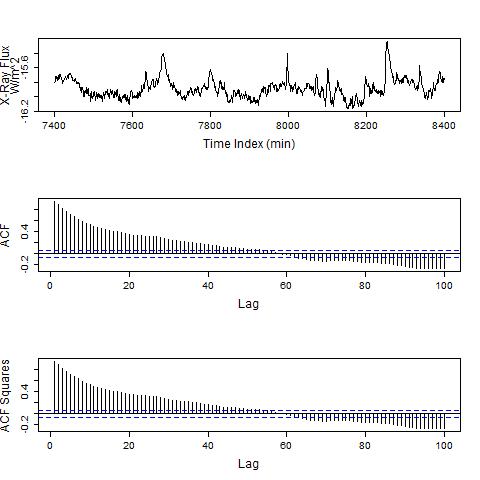}
\caption{Times series (top), ACF of time series and ACF of squared time series (bottom) of log transformed data for July 2017 data}
\label{fig:avgsolar_state2_jul_acfplot}
\end{figure*}

To make the dataset stationary, we take the log-transform of the data and use a shorter length time series that makes it more stationary. The Augmented Dickey-Fuller test \cite{said1984testing} rejects the null hypothesis for lack of unit-root stationarity with a p-value of 0.1 and fails to reject null hypothesis of trend stationarity using the KPSS test \cite{kwiatkowski1992testing}. The ACF plots in Figure \ref{fig:avgsolar_state2_jul_acfplot} for the log-transformed data and the squared log-transformed data indicate evidence of long memory.

We fit the log-transformed samples of time series with an ARTFIMA$(1,d,\lambda,1)$ model. The long memory parameter estimate $d>0.5$, which indicates lack of stationarity \cite{bhansali2001estimation} when using a ARFIMA model.
We use the {\tt artfima} R package to fit the ARTFIMA and ARFIMA models.
To check if the data follows heavy tails, we estimate the stability parameter $\alpha$ using a quantile based method called McCulloch estimation \cite{mcculloch1986simple}. The value of the $\alpha$ estimate for the solar flare data for the month of July is $1.639$ which indicates a stable time series with infinite second moment. Hence, we use the Whittle estimator to estimate the parameters of the ARTFIMA model. Table \ref{tab:stable_param_est} shows the parameter estimate for the month of July using ARTFIMA model. We shall mention that one cannot use the ARFIMA model for this dataset since the ARFIMA model requires to have $d<1-1/\alpha=0.389$ for $\alpha=1.639$.

\begin{table}[htpb]
\centering
\caption{Whittle estimates of $d, \lambda, \theta_1,\phi_1$ for July 2017 data.}
\label{tab:stable_param_est}
\begin{tabular}{|c|c|}
\hline
Parameter & ARTFIMA Parameter Estimates\\\hline
$d$ & 0.611  \\
$\lambda$ & 0.026 \\
$\phi_1$ &  0.652  \\
$\theta_1$ & 0.225 \\
\hline
\end{tabular}
\end{table}

\begin{figure*}[htpb]
\centering
\includegraphics[width=0.48\linewidth]{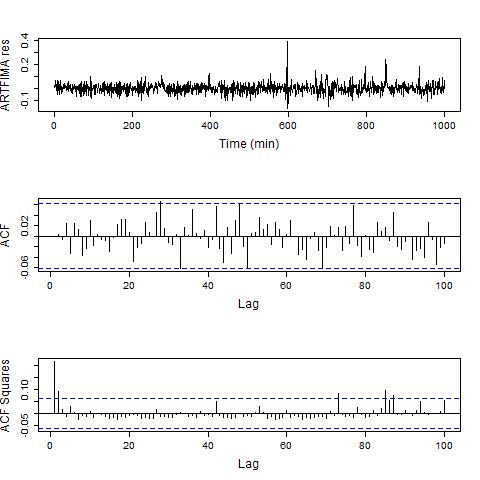}
\includegraphics[width=0.48\linewidth]{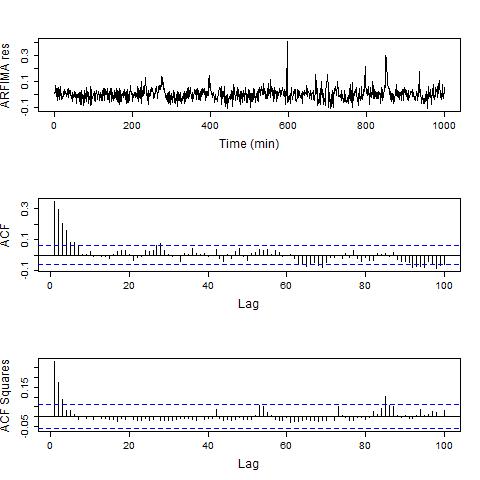}
\caption{Comparison of residuals and ACF plots of residuals and squared residuals between the ARTFIMA (left) and ARFIMA model fitted  with the log transformed data for July 2017 data}
\label{fig:avgsolar_state2_jul_tsplot}
\end{figure*}


We can verify the independence of residuals and square residuals for the ARTFIMA model and compare it against ARFIMA model. The lack of fit of residuals in the ARFIMA model is obvious from Figure \ref{fig:avgsolar_state2_jul_tsplot} since the ACF of the residuals lies outside the 95\%  confidence interval.  We also verify the lack of independence of ARFIMA residuals using a Ljung box test \cite{ljung1978measure} with a $p$-value 0.001 indicating reject the null hypothesis of i.i.d residuals. On the other hand, ARTFIMA residuals result in a p-value of 0.965, indicating strong evidence for i.i.d residuals.
%

\section{Appendix}\label{proofs}

{\it Proof of Proposition \ref{th:marep}:}
Define $A_{\lambda}{B}:=\Delta^{-d,\lambda}\ \Theta_q(B){\Phi_p(B)}^{-1}$ and write ${\Theta_q(z)}{\Phi_p(z)}^{-1}=\sum_{j=0}^{\infty}b(j)z^{j}$ for $|z|\leq 1$. Then
\begin{equation*}
A_{\lambda}(z)=(1-e^{-\lambda}z)^{-d}\ {\Theta_q(z)}{\Phi_p(z)}^{-1}=\Big(\sum_{i=0}^{\infty}\omega_{-d,\lambda}(i)z^{i}\Big)
\Big(\sum_{s=0}^{\infty}b(s)z^{s}\Big)=\sum_{j=0}^{\infty}a_{-d,\lambda}(j)z^{j},
\end{equation*}
where
\begin{equation}\label{qe:aj defn}
a_{-d,\lambda}(j)=\sum_{s=0}^{j}\omega_{-d,\lambda}(s)b(j-s)
\end{equation}
for $j\geq 0$. Since $X(t)$ satisfies $\Phi(B)X(t) = \Theta(B)\Delta^{-d,\lambda} Z(t)$, we can write
\begin{equation}\label{eq:artfima causal}
\begin{split}
X(t)=\Delta^{-d,\lambda}\frac{\Theta_q(B)}{\Phi_p(B)}Z(t)=
\left(\sum_{j=0}^{\infty}a_{-d,\lambda}(j)B^{j}\right)Z(t)&=\sum_{j=0}^{\infty}a_{-d,\lambda}(j)Z(t-j)\\
&=:X_{p,d,\lambda,q},
\end{split}
\end{equation}
where $a_{-d,\lambda}(j)$ is given by \eqref{qe:aj defn} and hence this proves \eqref{movingaveragerep}. Next, in order to show the series in \eqref{movingaveragerep} converges a.s. and in $L^{\nu}$ for any $\nu<\alpha$, we need to show that $\sum_{j=0}^{\infty}|a_{-d,\lambda}(j)|^{\nu}<\infty$ for some $\nu<\alpha$. Consider the case $1\leq \nu<\alpha<2$ and write
\begin{equation}\label{sumaj}
\sum_{j=0}^{\infty}\big|a_{-d,\lambda}(j)\big|=\sum_{j=0}^{\infty}\big|\sum_{s=0}^{j}\omega_{-d,\lambda}(s)b_{j-s}\big|
=\sum_{j=0}^{\infty}|(\omega_{-d,\lambda}* b)(j)|^{\nu}=\|\omega_{-d,\lambda}* b\|_{\nu}^{\nu},
\end{equation}
where $(\omega_{-d,\lambda}* b)(j)=\sum_{s=0}^{j}\omega_{-d,\lambda}(s)b_{j-s}$. Under {\bf Assumption 2}, $\big|{\Theta_q(z)}/{\Phi_p(z)}\big|<\infty$, for $|z|\leq 1+\varepsilon$, and the convergence of the series  ${\Theta_q(z)}/{\Phi_p(z)}$ implies that $|b_j|\leq C(1+\varepsilon)^{-j}$ for $j\geq 0$ (e.g., see \cite[Theorem 7.2.3]{giraitis2012book} or \cite[Theorem 3.1.1]{brockwell21time}) and hence $b(j)\in L^{1}$. On the other side,
\begin{equation}
\sum_{j=0}^{\infty}|\omega_{-d,\lambda}(j)|^{\nu}= \sum_{j=0}^{M}|\omega_{-d,\lambda}(j)|^{\nu}
+\Gamma(\nu)^{-d}\sum_{j=M+1}^{\infty}\big(j^{d-1}e^{-\lambda j}\big)^{\nu}<\infty
\end{equation}
and hence $\omega_{-d,\lambda}(j)\in L^{\nu}$. Now, by applying Young inequality we have
\begin{equation}
\|\omega_{-d,\lambda}* b\|_{\nu}\leq \|b\|_{1} \|\omega_{-d,\lambda}\|_{\nu}<\infty
\end{equation}
and this shows $\sum_{j=0}^{\infty}|a_{-d,\lambda}(j)|^{\nu}<\infty$ for some $1\leq \nu<\alpha<2$. Consequently, the series in \eqref{movingaveragerep} converges a.s. and in $L^{\nu}$ for any $1\leq \nu<\alpha<2$. The proof for the case
$0\leq \nu<\alpha<1$ is similar and the only difference is to show that
\begin{equation}\label{Youngnuless1}
\|\omega_{-d,\lambda}* b\|_{\nu}\leq \|b\|_{\nu} \|\omega_{-d,\lambda}\|_{\nu}<\infty.
\end{equation}
Note that since $|b_j|\leq C(1+\varepsilon)^{-j}$ for $j\geq 0$ then $|b_j|^{\nu}\leq C(1+\varepsilon)^{-\nu j}$ for $j\geq 0$
implying that $b(j)\in L^{\nu}$. This proves Consequently, the series in \eqref{movingaveragerep} converges a.s. and in $L^{\nu}$ for any $0\leq \nu<\alpha<1$. Therefore Minkowski's inequality implies
\begin{equation}
\mathbb{E}|X_{p,d,\lambda,q}(t)|^{\nu}\leq 2\mathbb{E}|Z(0)|^{\nu}\sum_{j=0}^{\infty}|a_{-d,\lambda}(j)|^{\nu}<\infty
\end{equation}
for $0<\nu<\alpha<2$. The proof of part (a) is completed now.

To prove part (b), let $C_{d,\lambda}(B):=\Delta^{d,\lambda}\ {\Phi_p(B)}/{\Theta_q(B)}$. Write ${\Phi(z)}/{\Theta(z)}=\sum_{j=0}^{\infty}c(j)z^{j}$ for $|z|\leq 1$ so that
\begin{equation*}
C_{d,\lambda}(z)=(1-e^{-\lambda}z)^{d}\ \frac{\Phi_p(z)}{\Theta_q(z)}=\Big(\sum_{i=0}^{\infty}\omega_{d,\lambda}(i)z^{i}\Big)
\Big(\sum_{s=0}^{\infty}c(s)z^{s}\Big)=\sum_{j=0}^{\infty}c_{d,\lambda}(j)z^{j},
\end{equation*}
where
\begin{equation}\label{eq:calphaj defn}
c_{d,\lambda}(j)=\sum_{s=0}^{j}\omega_{d,\lambda}(s)c(j-s)
\end{equation}
for $j\geq 0$. Since $Z(t)$ satisfies $\Phi(B)X(t) = \Theta(B)\Delta^{-d,\lambda} Z(t)$, we can write
\begin{equation}\label{eq:artfimainv}
Z(t)=\Delta^{d,\lambda}\frac{\Phi_p(B)}{\Theta_q(B)}Z(t)=
\left(\sum_{j=0}^{\infty}c_{d,\lambda}(j)B^{j}\right)X(t)=\sum_{j=0}^{\infty}c_{d,\lambda}(j)X(t-j)\\
\end{equation}
where $c_{d,\lambda}(j)$ is given by \eqref{eq:calphaj defn} and hence this proves \eqref{invertrep}. Next, in order to show the series in \eqref{invertrep} converges a.s. and in $L^{\nu}$ for any $\nu<\alpha$, one need to verify that $\sum_{j=0}^{\infty}|c_{d,\lambda}(j)|^{\nu}<\infty$ for some $\nu<\alpha$. But, the proof is similar to part (a) and hence we omit the details. The proofs of part (b) and Proposition \ref{th:marep} is completed now.
\hfill $\Box$

{\it Proof of Theorem \ref{codifferenceless1}:}
Let
\begin{equation}\label{codifferencedefinition}
\begin{split}
\tau_{d,\lambda}(n):&=\tau(X_{d,\lambda}(0),X_{d,\lambda}(n))\\
&=\sum_{j=0}^{\infty}\big[|\omega_{-d,\lambda}(j)|^{\alpha}+|\omega_{-d,\lambda}(j+n)|^{\alpha}-|\omega_{-d,\lambda}(j) - \omega_{-d,\lambda}(j+n)|^{\alpha}\big]\\
&:=I_{1}+ I_{2},
\end{split}
\end{equation}
where $I_{1}=\sum_{j=0}^{\infty}|\omega_{-d,\lambda}(j+n)|^{\alpha}$ and $I_2=\sum_{j=0}^{\infty}|\omega_{-d,\lambda}(j)|^{\alpha}-|\omega_{-d,\lambda}(j) - \omega_{-d,\lambda}(j+n)|^{\alpha}$. Since $\omega_{-d,\lambda}(j)\sim \frac{1}{\Gamma(d)}e^{-\lambda j}j^{d-1}$ as $j\to\infty$, we may work with the asymptotic form of $\omega_{-d,\lambda}(j)$.
For any $j>0$,
\begin{equation}\label{I1step1}
\begin{split}
e^{\lambda\alpha n} n^{-\alpha(d-1)}|\omega_{-d,\lambda}(j)|^{\alpha}&=C e^{\lambda\alpha n} n^{-\alpha(d-1)} |e^{-\lambda\alpha (n+j)}(n+j)^{d-1}|\\
&=e^{-\lambda\alpha j}\Big(\frac{j+n}{n}\Big)^{\alpha(d-1)}\to e^{-\alpha\lambda j}\ {\rm as}\ n\to\infty.
\end{split}
\end{equation}
We note that
\begin{equation}\label{I1step2}
\begin{split}
\sup_{n>1}\Big( \Big|e^{\lambda\alpha n} n^{-\alpha(d-1)} \omega_{-d,\lambda}(n+j)\Big|  \Big)
&=C\sup_{n>1}\Big( \Big|e^{-\lambda\alpha j} \Big(\frac{j+n}{n}\Big)^{\alpha(d-1)}\Big| \Big)\\
&\leq \begin{cases}
e^{-\lambda\alpha j}, &\alpha(d-1)\leq 0, \\
e^{-\lambda\alpha j} (1+j)^{\alpha(d-1)},&  \alpha(d-1)>0,
\end{cases}
\end{split}
\end{equation}
which belongs to $L^{1}(0,\infty)$. Now, using \eqref{I1step1}, \eqref{I1step2}, the dominated convergence theorem implies that
\begin{equation}\label{I1step3}
\begin{split}
e^{\lambda\alpha n}n^{-\alpha(d-1)}I_1&=C e^{\lambda\alpha n}n^{-\alpha(d-1)}\sum_{j=0}^{\infty}(j+n)^{\alpha(d-1)}e^{-\lambda\alpha(j+n)}\\
&\qquad\qquad\qquad\to \sum_{j=0}^{\infty}e^{-\lambda\alpha j}=\frac{1}{1-e^{-\lambda\alpha}},\ {\rm as}\ n\to\infty.
\end{split}
\end{equation}
Next, we show $e^{\lambda\alpha n}n^{-\alpha(d-1)}I_2\to 0$ as $n\to\infty$. For each $j>0$,
\begin{equation}\label{I2step1}
\begin{split}
&e^{\lambda\alpha n} n^{-\alpha(d-1)}\Big[ |\omega_{-d,\lambda}(j)|^{\alpha}-|\omega_{-d,\lambda}(j) - \omega_{-d,\lambda}(j+n)|^{\alpha}\Big]\\
&=\Big| \Big(\frac{j}{n}\Big)^{\alpha(d-1)} e^{-\lambda\alpha(j- n/\alpha)} \Big|
- \Big| \Big(\frac{j}{n}\Big)^{(d-1)} e^{-\lambda(j-n)} - \Big(\frac{n+j}{n}\Big)^{(d-1)} e^{-\lambda j}  \Big|^{\alpha}\\
&=:|a_n + b_n|^{\alpha}- |b_n|^{\alpha},
\end{split}
\end{equation}
where $a_n=-\Big(\frac{n+j}{n}\Big)^{(d-1)} e^{-\lambda j}$ and $b_n=\Big(\frac{j}{n}\Big)^{(d-1)} e^{-\lambda(j-n)}$. It is obvious that
$a_n\to -e^{-\lambda j}$ and $b_n\to\infty$ as $n\to\infty$. Then using $|a_n+b_n|^{\alpha}-|b_n|^{\alpha}\to 0$ as $n\to\infty$
since $0<\alpha\leq 1$ and hence
\begin{equation}\label{I2step2}
e^{\lambda\alpha n} n^{-\alpha(d-1)}\Big[ |\omega_{-d,\lambda}(j)|^{\alpha}-|\omega_{-d,\lambda}(j) - \omega_{-d,\lambda}(j+n)|^{\alpha}\Big]
\to 0
\end{equation}
as $n\to\infty$. Now, using the fact that $\Big| |a|^{\alpha}-|b|^{\alpha} \Big|\leq |a-b|^{\alpha}$, for $a,b \in\mathbb{R}$ and $0<\alpha\leq 1$, we have
\begin{equation}\label{I2step3}
\Big| |j^{d-1}e^{-\lambda j}|^{\alpha} - |j^{d-1}e^{-\lambda j} - (n+j)^{d-1}e^{-\lambda (n+j)}|^{\alpha}   \Big|
\leq |(n+j)^{d-1}e^{-\lambda(n+j)}|^{\alpha}
\end{equation}
and hence
\begin{equation}\label{I2step4}
\begin{split}
&\sup_{n>1}\Big| e^{\lambda\alpha n} n^{-\alpha(d-1)}\Big[ |\omega_{-d,\lambda}(j)|^{\alpha}-|\omega_{-d,\lambda}(j) - \omega_{-d,\lambda}(j+n)|^{\alpha}\Big]\Big|\leq
\sup_{n>1}\Big| e^{\lambda\alpha n} n^{-\alpha(d-1)} \omega_{-d,\lambda}(j)^{\alpha} \Big|\\
&\qquad\qquad\qquad\qquad\qquad\qquad\qquad\qquad\leq \begin{cases}
e^{-\lambda\alpha j}, &\alpha(d-1)\leq 0, \\
e^{-\lambda\alpha j} (1+j)^{\alpha(d-1)},&  \alpha(d-1)>0,
\end{cases}
\end{split}
\end{equation}
From \eqref{I2step1}-\eqref{I2step4}, the dominated convergence theorem implies that
\begin{equation}\label{I2step5}
e^{\lambda\alpha n}n^{-\alpha(d-1)}I_2\to 0\ {\rm as}\ n\to\infty.
\end{equation}
Finally, \eqref{I1step3} and \eqref{I2step5} together yield
\begin{equation}
\lim_{n\to\infty}\frac{\tau_{d,\lambda}(n)}{e^{-\lambda\alpha n}n^{\alpha(d-1)}}=\Gamma(d)^{-\alpha}(1-e^{-\lambda \alpha})^{-1}
\end{equation}
for $0<\alpha<1$ and $d\in\mathbb{R}-\mathbb{N}_{-}$. The proof is completed now.
\hfill $\Box$

{\it Proof of Theorem \ref{codifferencegrq1}:}
Recall from the proof of Theorem \ref{codifferenceless1} that
$I_{1}=\sum_{j=0}^{\infty}|\omega_{-d,\lambda}(j+n)|^{\alpha}$ and $I_2=\sum_{j=0}^{\infty}|\omega_{-d,\lambda}(j)|^{\alpha}-|\omega_{-d,\lambda}(j) - \omega_{-d,\lambda}(j+n)|^{\alpha}$. We may work with the asymptotic form of $\omega_{-d,\lambda}(j)$ as we did in proof of Theorem \ref{codifferenceless1}.
For any $j>0$,
\begin{equation}\label{I12step1}
\begin{split}
e^{\lambda n} n^{-(d-1)}|\omega_{-d,\lambda}(j)|^{\alpha}&=C e^{\lambda n} n^{-(d-1)} |e^{-\lambda\alpha (n+j)}(n+j)^{d-1}|\\
&=e^{-\lambda\alpha j} e^{-\lambda n(\alpha-1)} \Big(\frac{j+n}{n^{1/\alpha}}\Big)^{\alpha(d-1)}\to 0\ {\rm as}\ n\to\infty.
\end{split}
\end{equation}
since $1<\alpha\leq 2$.
We note that
\begin{equation}\label{I12step2}
\begin{split}
\sup_{n>1}\Big( \Big|e^{\lambda n} n^{-(d-1)} \omega_{-d,\lambda}(n+j)\Big|  \Big)
&=C\sup_{n>1}\Big( \Big| e^{-\lambda\alpha j} e^{-\lambda n(\alpha-1)} \Big(\frac{j+n}{n^{1/\alpha}}\Big)^{\alpha(d-1)}\Big| \Big)\\
&\leq \begin{cases}
e^{-\lambda\alpha j}, &d-1\leq 0, \\
e^{-\lambda\alpha j} (1+j)^{\alpha(d-1)},&  d-1>0,
\end{cases}
\end{split}
\end{equation}
which belongs to $L^{1}(0,\infty)$. Now, using \eqref{I12step1}, \eqref{I12step2}, the dominated convergence theorem implies that
\begin{equation}\label{I12step3}
\begin{split}
e^{\lambda n}n^{(d-1)}I_1&=C e^{\lambda n}n^{(d-1)}\sum_{j=0}^{\infty}(j+n)^{\alpha(d-1)}e^{-\lambda\alpha(j+n)}\\
&\qquad\qquad\qquad\to 0,\ {\rm as}\ n\to\infty.
\end{split}
\end{equation}
Next, we show $e^{\lambda n}n^{-(d-1)}I_2\to \alpha e^{-\lambda\alpha j} j^{(d-1)(\alpha-1)}$ as $n\to\infty$. For each $j>0$,
\begin{equation}\label{I22step1}
\begin{split}
&e^{\lambda n} n^{-(d-1)}\Big[ |\omega_{-d,\lambda}(j)|^{\alpha}-|\omega_{-d,\lambda}(j) - \omega_{-d,\lambda}(j+n)|^{\alpha}\Big]\\
&=\Big| \Big(\frac{j}{n^{1/\alpha}}\Big)^{\alpha(d-1)} e^{-\lambda\alpha(j-n)} \Big|
- \Big| \Big(\frac{j}{n^{1/\alpha}}\Big)^{(d-1)} e^{-\lambda(j-n/\alpha)} - \Big(\frac{n+j}{n^{1/\alpha}}\Big)^{(d-1)} e^{-\lambda j}  e^{-\lambda n(1-1/\alpha)}\Big|^{\alpha}\\
&=:|a_n|^{\alpha}- |a_n-b_n|^{\alpha},
\end{split}
\end{equation}
where $a_n=\Big(\frac{j}{n^{1/\alpha}}\Big)^{(d-1)} e^{-\lambda(j-n/\alpha)}$ and $b_n=\Big(\frac{n+j}{n^{1/\alpha}}\Big)^{(d-1)} e^{-\lambda j}e^{-\lambda n(1-1/\alpha)}$. It is obvious that
$a_n\to \infty$ and $b_n\to 0$ as $n\to\infty$. Then using $|a_n|^{\alpha}-|a_n-b_n|^{\alpha}\sim
\alpha|b_n||a_n|^{\alpha-1}$, as $n\to\infty$, we get
\begin{equation}\label{I22step2}
e^{\lambda n} n^{-(d-1)}\Big[ |\omega_{-d,\lambda}(j)|^{\alpha}-|\omega_{-d,\lambda}(j) - \omega_{-d,\lambda}(j+n)|^{\alpha}\Big]
\sim \alpha e^{-\alpha\lambda j}\Big(\frac{n+j}{nj}\Big)^{d-1} j^{\alpha(d-1)}
\end{equation}
consequently,
\begin{equation}\label{I22step3}
e^{\lambda n} n^{-(d-1)}\Big[ |\omega_{-d,\lambda}(j)|^{\alpha}-|\omega_{-d,\lambda}(j) - \omega_{-d,\lambda}(j+n)|^{\alpha}\Big]
\to  \alpha e^{-\alpha\lambda j} j^{(\alpha-1)(d-1)}.
\end{equation}
Now, using the fact that $\Big| |a_n-b_n|^{\alpha}-|a_n|^{\alpha} \Big|\leq b_n^{\alpha} + \alpha b_n {a_n}^{\alpha-1}$, for $a,b\geq 0$ and $1<\alpha\leq 2$, we have
\begin{equation}\label{I22step4}
\begin{split}
&\sup_{n>1}\Big| e^{\lambda n} n^{-(d-1)}\Big[ |\omega_{-d,\lambda}(j)|^{\alpha}-|\omega_{-d,\lambda}(j) - \omega_{-d,\lambda}(j+n)|^{\alpha}\Big]\Big|\leq
\sup_{n>1} b_n^{\alpha} + \alpha \sup_{n\geq 1}a_n b_n^{\alpha-1}\\
&\qquad\qquad\qquad\qquad\qquad\qquad\leq \begin{cases}
e^{-\lambda\alpha j}\big[ 1+ \alpha(j+1)^{d-1}j^{(d-1)(\alpha-1)} \big], &d-1 \leq 0, \\
e^{-\lambda\alpha j}\big[ \alpha(j+1)^{\alpha(d-1)} \alpha j^{(d-1)} \big],&  (d-1)>0,
\end{cases}
\end{split}
\end{equation}
which belongs to $L^{1}$. From \eqref{I22step1}-\eqref{I22step4}, the dominated convergence theorem implies that
\begin{equation}\label{I22step5}
e^{\lambda n}n^{-\alpha(d-1)}I_2\to \alpha \sum_{j=0}^{\infty} e^{-\lambda\alpha j} j^{(d-1)(\alpha-1)}\ {\rm as}\ n\to\infty.
\end{equation}
Finally, \eqref{I12step3} and \eqref{I22step5} together yield
\begin{equation}
\lim_{n\to\infty}\frac{\tau_{d,\lambda}(n)}{e^{-\lambda n}n^{(d-1)}}=(\Gamma(d))^{-\alpha}\sum_{j=0}^{\infty}\alpha e^{-\lambda\alpha j}\omega_{-d}^{\alpha-1}(j)
\end{equation}
for $1<\alpha<2$ and $d\in\mathbb{R}-\mathbb{N}_{-}$. The proof is completed now.
\hfill $\Box$

{\it Proof of Corollary \ref{semilong}:}
Theorems \ref{codifferenceless1} and \ref{codifferencegrq1} imply that $\sum_{j=0}^{\infty}|\tau_{d,\lambda}(n)|<\infty$ for any $d\in\mathbb{R}-\mathbb{N}_{-}$ and $\alpha\in (0,2)$ and hence the statement holds.
\hfill $\Box$

{\it proof of Lemma \ref{lemma1consistency}:}
We use the asymptotic behavior of $\omega_{-d,\lambda}(j)$ to conclude the statement of the Lemma as follows:
\begin{equation*}
\sum_{j=0}^\infty |\omega_{-d,\lambda}(j)|^\delta\ |j|= \sum_{j=0}^M |\omega_{-d,\lambda}(j)|^\delta\ |j|+
C\sum_{j=M+1}^\infty |j^{d-1}e^{-\lambda j}|^\delta\ |j|<\infty
\end{equation*}
for any $d\in\mathbb{R}-\mathbb{Z}_{-}$ and this completes the proof.
\hfill $\Box$

{\it Proof of Theorem \ref{thm:asymptotic_dis_covariance}:}
The proof follows by \cite[Theorem 13.3.1]{brockwell21time} if we can verify the claim that $\sum_{j=0}^{\infty}|j| |\omega_{-d,\lambda}(j)|^{\delta}<\infty$ for $\delta\in (0,\alpha)\cap [0,1]$. But this claim proved in Lemma \eqref{lemma2consistency}
and this completes the proof.
\hfill $\Box$

{\it Proof of Lemma \ref{lemma2consistency}:}
We shall only prove \eqref{part2_lemma2consistency}. The proof of \eqref{part1_lemma2consistency} is analogous. Let
\[q_m(\omega,\beta) =\frac{1}{m} \sum_{j=0}^{m-1}\sum_{|k|\leq j} r_k e^{-i\omega k} = \sum_{|k|< m} \Big( 1 - \frac{|k|}{m} \Big)r_k e^{-i\omega k},\]
where
\[r_k = \frac{1}{2\pi}\int_{-\pi}^{\pi}e^{i\omega k} \Big(g^{\delta}_{p,d,\lambda,q}(\omega)\Big)^{-1}\ d\omega.\]
We note that $q_m(\omega,\beta)\geq 0$ since the $\{r_k\}$ are non-negative definiteness. We also see that $\big(g^{\delta}_{p,d,\lambda,q}\big)^{-1}$ is a uniformly continuous function with respect to $(\omega,\beta)$ on $\bar{E}$. Now, \cite[Theorem 2.11.1]{brockwell21time} implies that
$q_m(\omega,\beta)$ converges uniformly to $\Big(g^{\delta}_{p,d,\lambda,q}\Big)^{-1}$ on $\bar{E}$. This means for any
$\epsilon>0$, then there exists an $m\in \mathbb{N}$ such that
\[\big| q_m(\omega,\beta)-\big(g^{\delta}_{p,d,\lambda,q}(\omega,\beta)\big)^{-1}\big|<\frac{\epsilon}{4\pi}.\]
The rest of the proof is similar to \cite[Lemma 6.1]{mikosch1995parameter} since $\tilde{\gamma}_{n,X}\to \tilde{\gamma}(h)$ by Theorem \ref{thm:asymptotic_dis_covariance}. The proof of the Lemma is complete now.
\hfill $\Box$

{\it Proof of Lemma \ref{lemma3consistency}:}

Let $|z|<1$ and $\beta\in E$. Recall from the proof of Proposition \ref{th:marep},
\begin{equation*}
\begin{split}
A_{\beta}(z)&=A_{p,d,\lambda,q}(z)=(1-e^{-\lambda}z)^{-d}\ {\Theta_q(z)}{\Phi_p(z)}^{-1}\\
&=\Big(\sum_{i=0}^{\infty}\omega_{-d,\lambda}(i)z^{i}\Big)
\Big(\sum_{s=0}^{\infty}b(s)z^{s}\Big)=\sum_{j=0}^{\infty}a_{-d,\lambda}(j)z^{j},
\end{split}
\end{equation*}
where $a_{p,-d,\lambda,q}(j)=a_{-d,\lambda}(j)$ is given by \eqref{qe:aj defn}. We also defined,
\begin{equation*}
\begin{split}
A_{\beta}(z)&=C_{p,d,\lambda,q}(z)=(1-e^{-\lambda}z)^{d}\ \frac{\Phi_p(z)}{\Theta_q(z)}\\
&=\Big(\sum_{i=0}^{\infty}\omega_{d,\lambda}(i)z^{i}\Big)
\Big(\sum_{s=0}^{\infty}c(s)z^{s}\Big)=\sum_{j=0}^{\infty}c_{d,\lambda}(j)z^{j},
\end{split}
\end{equation*}
where $c_{p,d,\lambda,q}(j)=c_{d,\lambda}(j)$ is given by \eqref{eq:calphaj defn}. Let $\xi(n)$ be a sequence of Gaussian random variables with zero mean and unit variance, and $X_{\beta_1}(n)=\sum_{j=0}^{\infty}a_{-d_1,\lambda_1}(j)\xi(n-j)$ be a Gaussian ARTFIMA$(p_1,d_1,\lambda_1,q_1)$. By mimicking a similar argument in \cite[Section 13.2]{brockwell21time}, it can be shown that
${\rm Var}\big(X_{\beta_1}(n+1)-\sum_{j=0}^{\infty}{\theta(j)}X_{\beta_1}(n-j) \big)$ will be minimized if and only if $\theta(j)=-c_{p,d_1,\lambda_1,q}(j)$ and the smallest value of the variance is one. Now, for $\beta_1\neq \beta_2$, we have $C_{\beta_1}(z)\neq C_{\beta_2}(z)$, and hence ${\rm Var}\big(X_{\beta_1}(n+1)+\sum_{j=0}^{\infty}{c_{p,d_1,\lambda_1,q}(j)}X_{\beta_1}(n-j) \big)>1$. Therefore
\begin{equation*}
\begin{split}
\frac{1}{2\pi}\int_{-\pi}^{\pi}\frac{g_{p_1,d_1,\lambda_1,q_1}(\omega,\beta_1)}{g_{p_2,d_2,\lambda_2,q_2}(\omega,\beta_2)}
&=\frac{1}{2\pi}\int_{-\pi}^{\pi}\Big| \frac{A_{d_1,\lambda_1}(e^{-i\omega})}{A_{d_2,\lambda_2}(e^{-i\omega})}\Big|^2\ d\omega \\
&=\frac{1}{2\pi}\int_{-\pi}^{\pi}\Big| C_{d_2,\lambda_2}(e^{-i\omega}) A_{d_1,\lambda_1}(e^{-i\omega}) \Big|^2\ d\omega\\
&=\sum_{j=0}^{\infty}\Big|\sum_{s=0}^{j}c_{d_2,\lambda}(s)a_{-d_1,\lambda_1}(j-s)\Big|^2\\
&={\rm Var}\big(X_{\beta_1}(n+1)-\sum_{j=0}^{\infty}{\theta(j)}X_{\beta_1}(n-j) \big)>1
\end{split}
\end{equation*}
and this completes the proof.
\hfill $\Box$

{\it Proof of Theorem \ref{thm:consistency}:}
The proof follows by Lemmas \ref{lemma2consistency} and \ref{lemma3consistency} and mimicking a similar arguments as for proving
\cite[Theorem 1.1.]{kokoszka1996parameter}.

{\it Proof of Theorem \ref{thm:asymptotic_dis_parameters}:}
The proofs follow a similar path of the proof of \cite{mikosch1995parameter}, and hence we omit the details.

\bibliographystyle{abbrv}
\bibliography{paper1}
\end{document}